\documentclass[12pt,preprint]{aastex}
\usepackage{CJK}
\usepackage{multirow}
\usepackage{color}

\newcommand{\bdv}[1]{\mbox{\boldmath$#1$}}

\slugcomment{}
\shortauthors{Zhu et al.}
\begin{document}

\title{\emph{Spitzer} as Microlens Parallax Satellite:\\Mass and Distance Measurements of Binary Lens System OGLE-2014-BLG-1050L}
\begin{CJK*}{UTF8}{gkai}

\author{
Wei~Zhu~(祝伟)~$^{1}$,
A.~Udalski~$^{2,O}$,
A.~Gould~$^{1,U}$,
M.~Dominik~$^{3,}$\altaffilmark{\#},
V.~Bozza~$^{4,5}$,
C.~Han~$^{6}$,
J.~C.~Yee~$^{7,U,}$\altaffilmark{*},
S.~Calchi~Novati~$^{8,4,9,}$\altaffilmark{@},
C.~A.~Beichman~$^{8}$,
S.~Carey~$^{10}$
\vspace{-0.3cm}
\begin{center}
    and
\end{center}
\vspace{-0.3cm}
R.~Poleski~$^{1,2}$,
J.~Skowron~$^{2}$,
S.~Koz{\l}owski~$^{2}$,
P.~Mr{\'o}z~$^{2}$,
P.~Pietrukowicz~$^{2}$,
G.~Pietrzy{\'n}ski~$^{2,11}$
M.~K.~Szyma{\'n}ski~$^{2}$,
I.~Soszy{\'n}ski~$^{2}$,
K.~Ulaczyk~$^{2}$,
{\L}.~Wyrzykowski~$^{2,12}$
\vspace{-0.3cm}
\begin{center}
(The OGLE collaboration)
\end{center}
\vspace{-0.3cm}
B.~S.~Gaudi~$^{1}$,
R.~W.~Pogge~$^{1}$,
D.~L.~DePoy~$^{13}$
\vspace{-0.3cm}
\begin{center}
(The $\mu$FUN collaboration)
\end{center}
\vspace{-0.3cm}
Y.~K.~Jung~$^6$,
J.-Y.~Choi~$^6$,
K.-H.~Hwang~$^6$,
I.-G.~Shin~$^6$,
H.~Park~$^6$,
J.~Jeong~$^6$ \\
\normalsize{$^1$ Department of Astronomy, Ohio State University, 140 W. 18th Ave., Columbus, OH  43210, USA} \\
\normalsize{$^2$ Warsaw University Observatory, Al.~Ujazdowskie~4, 00-478~Warszawa, Poland} \\ 
\normalsize{$^3$ SUPA, School of Physics \& Astronomy, North Haugh, University of St Andrews, KY16 9SS, Scotland, UK} \\
\normalsize{$^4$ Dipartimento di Fisica ``E. R. Caianiello'', Universit\'a di Salerno, Via Giovanni Paolo II, 84084 Fisciano (SA), Italy} \\
\normalsize{$^5$ Istituto Nazionale di Fisica Nucleare, Sezione di Napoli, Via Cintia, 80126 Napoli, Italy} \\
\normalsize{$^6$ Department of Physics, Institute for Astrophysics, Chungbuk National University, Cheongju 371-763, Republic of Korea} \\
\normalsize{$^7$ Harvard-Smithsonian Center for Astrophysics, 60 Garden St., Cambridge, MA 02138, USA} \\
\normalsize{$^8$ NASA Exoplanet Science Institute, MS 100-22, California Institute of Technology, Pasadena, CA 91125, USA} \\
\normalsize{$^9$ Istituto Internazionale per gli Alti Studi Scientifici (IIASS), Via G. Pellegrino 19, 84019 Vietri Sul Mare (SA), Italy} \\
\normalsize{$^{10}$Spitzer Science Center, MS 220-6, California Institute of Technology,Pasadena, CA, USA} \\
\normalsize{$^{11}$ Universidad de Concepci{\'o}n, Departamento de Astronomia, Casilla 160--C, Concepci{\'o}n, Chile} \\
\normalsize{$^{12}$ Institute of Astronomy, University of Cambridge, Madingley Road, Cambridge CB3 0HA, UK} \\
\normalsize{$^{13}$Department of Physics and Astronomy, Texas A\& M University, College Station, TX 77843-4242, USA} 
\vspace{-0.3cm}
\begin{center}
\normalsize{$^{O}$(The OGLE collaboration)}\\
\normalsize{$^{U}$(The $\mu$FUN collaboration)}\\
\end{center}
\vspace{-0.3cm}
}

\altaffiltext{\#}{Royal Society University Research Fellow.}
\altaffiltext{*}{Sagan Fellow.}
\altaffiltext{@}{Sagan Visiting Fellow.}

\begin{abstract}
    We report the first mass and distance measurement of a caustic-crossing binary system OGLE-2014-BLG-1050L using the space-based microlens parallax method. \emph{Spitzer} captured the second caustic-crossing of the event, which occurred $\sim$10 days before that seen from Earth. Due to the coincidence that the source-lens relative motion was almost parallel to the direction of the binary-lens axis, the four-fold degeneracy, which was known before only to occur in single-lens events, persists in this case, leading to either a lower-mass (0.2 $M_\odot$ and 0.07 $M_\odot$) binary at $\sim$1.1 kpc or a higher-mass (0.9 $M_\odot$ and 0.35 $M_\odot$) binary at $\sim$3.5 kpc. However, the latter solution is strongly preferred for reasons including blending and lensing probability. OGLE-2014-BLG-1050L demonstrates the power of microlens parallax in probing stellar and substellar binaries.
\end{abstract}


\section{Introduction} \label{sec:introduction}

The detection of binary star systems depends on a combination of diverse observational techniques. For example, nearby wide binaries can be directly resolved by high-resolution imaging, while close binaries can be detected via eclipsing or spectroscopic methods. To get a full picture of the distributions of the mass ratios, masses and separations of binary systems, one also needs such a technique as microlensing to probe those binary systems that are difficult for other techniques, such as very low-mass binaries (i.e., brown dwarf binaries), dark binaries (e.g., binary black holes), and normal binaries with intermediate separations. For example, two brown dwarf binaries, OGLE-2009-BLG-151/MOA-2009-BLG-232 and OGLE-2011-BLG-0420, were detected via microlensing, with reported total masses of $0.025\ M_\odot$ and $0.034\ M_\odot$, respectively \citep{Choi:2013}.

The challenge faced by standard microlensing observations is to break the degeneracy between the mass of and the distance to the lens system, since these two physical parameters both enter a single observable quantity -- the timescale of the microlensing event, $t_{\rm E}$:
\[ t_{\rm E} = \frac{\theta_{\rm E}}{\mu_{\rm geo}};\quad \theta_{\rm E} = \sqrt{\kappa M_{\rm L} \pi_{\rm rel}};\quad \kappa \equiv \frac{4G}{c^2 \rm AU} \approx 8.14\frac{\rm mas}{M_\odot}.\]
Here $\theta_{\rm E}$ is the angular Einstein ring radius, $\mu_{\rm geo}$ is the geocentric lens-source relative proper motion, and 
\[ \pi_{\rm rel} \equiv \pi_{\rm L}-\pi_{\rm S}={\rm AU} \left(\frac{1}{D_{\rm L}} - \frac{1}{D_{\rm S}}\right) \]
is the lens-source relative parallax.

The so-called ``microlens parallax'' can in principle be used to meet this challenge \citep{Gould:1992}, since the measurement of the microlens-parallax amplitude, $\pi_{\rm E}\equiv \pi_{\rm rel}/\theta_{\rm E}$, directly leads to the determination of lens mass $M_{\rm L}$ and distance $D_{\rm L}={\rm AU}/\pi_{\rm L}$, by
\begin{equation} \label{eq:ml-dl}
    M_{\rm L} = \frac{\theta_{\rm E}}{\kappa \pi_{\rm E}};\quad
\pi_{\rm L} = \pi_{\rm E} \theta_{\rm E} + \pi_{\rm S}\ .
\end{equation}
To produce precise constraints on $M_{\rm L}$ and $D_{\rm L}$, one therefore needs precise measurements of both $\theta_{\rm E}$ and $\pi_{\rm E}$ (keeping in mind that $\pi_{\rm S}$ is usually known quite accurately). 

Two broad classes of methods have been proposed to measure the microlens parallax $\pi_{\rm E}$. The first is to obtain observations from a single platform that is being accelerated, which could be Earth \citep{Gould:1992}, or a satellite in low-Earth \citep{Honma:1999} or geosynchronous \citep{Gould:2013} orbit. This method has already produced $\sim$100 $\pi_{\rm E}$ measurements \citep[e.g.,][]{Alcock:1995,Poindexter:2005,Gaudi:2008}. However, it is strongly biased, in the case of binary-lens events, toward nearby lenses \citep{Choi:2013,Jung:2015}.
The other class of methods is to obtain observations from at least two well-separated observatories \citep{Refsdal:1966,HardyWalker:1995,Gould:1997}. In order to produce substantially different light curves observed by different observatories, which lead to precise measurement of $\pi_{\rm E}$, the required separation between observatories should be $\sim$ AU. For this reason, the so-called ``terrestrial microlens parallax'', i.e., using the ground-based observatories at different sites, only works in very rare cases \citep{Gould:2009,Yee:2009,GouldYee:2012}. The combination of ground-based observations and space-based observations from a satellite in solar orbit, called ``space-based microlens parallax'', has therefore been considered the only way to routinely measure the microlens parallax $\pi_{\rm E}$ for a substantial fraction of all microlensing events \citep{Refsdal:1966}.

There are also several methods proposed to measure the Einstein ring radius $\theta_{\rm E}$. In principle, $\theta_{\rm E}$ can be measured by resolving the light centroid of the multiple microlensed images \citep{Walker:1995}. However, such a method is presently inaccessible given the fact that the typical Galactic microlensing event has $\theta_{\rm E}\sim$ mas, although a future space telescope with ultra-precise astrometry may be able to achieve that \citep{GouldYee:2014}. The second method is to measure the geocentric lens-source relative proper motion, $\mu_{\rm geo}$, when the lens and source are well separated, typically a few years before or after the microlensing event occurs, since combining this measurement with $t_{\rm E}$ also yields $\theta_{\rm E}$ \citep{Alcock:2001}. However, since this method relies on light from both the lens and source, it does not work for a specific class of interesting events, i.e., events with very low-mass objects or stellar remnants as lenses. Nevertheless, this approach has been used several times to measure physical parameters of microlensing events \citep{Alcock:2001,Bennett:2006,Dong:2009}, and is likely to be used much more frequently in the future \citep{Gould:2015}. At present, the most widely exploited method of measuring $\theta_{\rm E}$ is using the finite source effect that is enabled when the source crosses or closely passes by the caustic structures of the lens system.

What makes binary star systems favorable targets for microlens parallax observations is that the finite source effect is often detected in binary lens events due to their relatively large caustics, producing precise measurement of $\theta_{\rm E}$ and, if the microlens parallax is also measured, the lens mass $M_{\rm L}$ and distance $D_{\rm L}$ \citep[also see][]{GraffGould:2002}. It is for this reason that special attention was paid to binary lens events when we were granted Director's Discretionary Time for a 100 hr pilot program to determine the feasibility of using \emph{Spitzer} as a parallax satellite, although the main objective was to measure lens masses in planetary events \citep{Udalski:2014}.

Here we report on the binary lens OGLE-2014-BLG-1050L, which is the second space-based parallax measurement of a binary lens \citep{Dong:2007} but the first such measurement for a caustic-crossing binary-lens event, or indeed, any caustic-crossing event. We give a summary of the observations from the ground and \emph{Spitzer} in Section~\ref{sec:observations}. The lightcurve modeling is demonstrated in Section~\ref{sec:fitting}; the source and blend characterizations and the derivations of physical parameters are presented in Section~\ref{sec:cmd} and Section~\ref{sec:physical}, respectively. In Section~\ref{sec:discussion} we present a discussion of our results.

\section{Observations} \label{sec:observations}

The source star for microlensing event OGLE-2014-BLG-1050 lies   toward the Galactic Bulge field with equatorial and Galactic coordinates 
(RA, Dec)$_{2000}=(17^{\rm h} 45^{\rm m} 07\fs 83,-22\arcdeg 54\arcmin 20\farcs 0)$ 
and $(l,b)_{2000}=(5\fdg 09,3\fdg 23)$, respectively. 
It therefore lies just $0\fdg49$ above the ecliptic plane. 

\subsection{Ground-based Observations}

The Optical Gravitational Lensing Experiment (OGLE) collaboration alerted the community to the new microlensing event OGLE-2014-BLG-1050 on 2014 June 6 (HJD$'$ = HJD$-2450000=6815.3$), just 35 hours before OGLE detected its first point on the caustic entrance at HJD$'$ = 6816.76, based on observations with the 1.4 deg$^2$ camera on its 1.3m Warsaw Telescope at the Las Campanas Observatory in Chile \citep{Udalski:2003}. 

OGLE-2014-BLG-1050 did not attract much attention when it was initially discovered. This situation changed after early modeling of online OGLE data revealed it was a binary-lens event.
Hence, in order to support the \emph{Spitzer} observations, the Microlensing Follow-Up Network ($\mu$FUN) started observing this event beginning HJD$'=6822$ using the 1.3 meter SMARTS telescope at Cerro Tololo InterAmerican Observatory (CTIO) (see Section~\ref{sec:spitzer}). Observations were taken only about once per night in the following few days. For this event, the most crucial task of follow-up teams is to capture the caustic exit with intensive observations so as to constrain $\theta_{\rm E}$ (see Section~\ref{sec:introduction}). However, it is very difficult to predict, based on modeling of the collected data, the exact time of caustic crossing, which normally lasts only a few hours. To achieve that goal, $\mu$FUN observed at one hour cadence and frequently reviewed these data during the night once the lightcurve began its slow rise toward the caustic exit. The cadence for OGLE-2014-BLG-1050 was increased to 10 per hour once the caustic exit was recognized.

\subsection{\emph{Spitzer} Observations} \label{sec:spitzer}

The general description of the \emph{Spitzer} observations of this pilot program is given in \citet{Udalski:2014}. In short, the program was organized into 2.6 hour windows roughly once per day from June 5 (HJD$'$ = 6814) to July 12 (HJD$'$ = 6851); observing targets had to be chosen and submitted by J.C.Y. and A.G. to the \emph{Spitzer} Science Center $\sim3$ days before the next observing run once per week.

For the particular case of OGLE-2014-BLG-1050, OGLE had not yet issued its alert when the decisions were made for the first week of observations on June 2 (HJD$'$ = 6811.1). 
On HJD$'$ = 6817.9, M.D. suggested OGLE-2014-BLG-1050 might be a binary-lens event, but this was too close to the deadline (6818.1) to modify the observing protocol prior to upload. During the following week two modelers (M.D. and V.B.) confirmed it as a binary-lens event, and therefore high priority was assigned to this event, leading to twice-per-day and once-per-day observations in weeks 3 and 4, respectively, and once or twice per day in week 5 until it moved beyond \emph{Spitzer}'s Sun-angle window at HJD$'$ = 6846.56.

\begin{figure}
\centering
\plotone{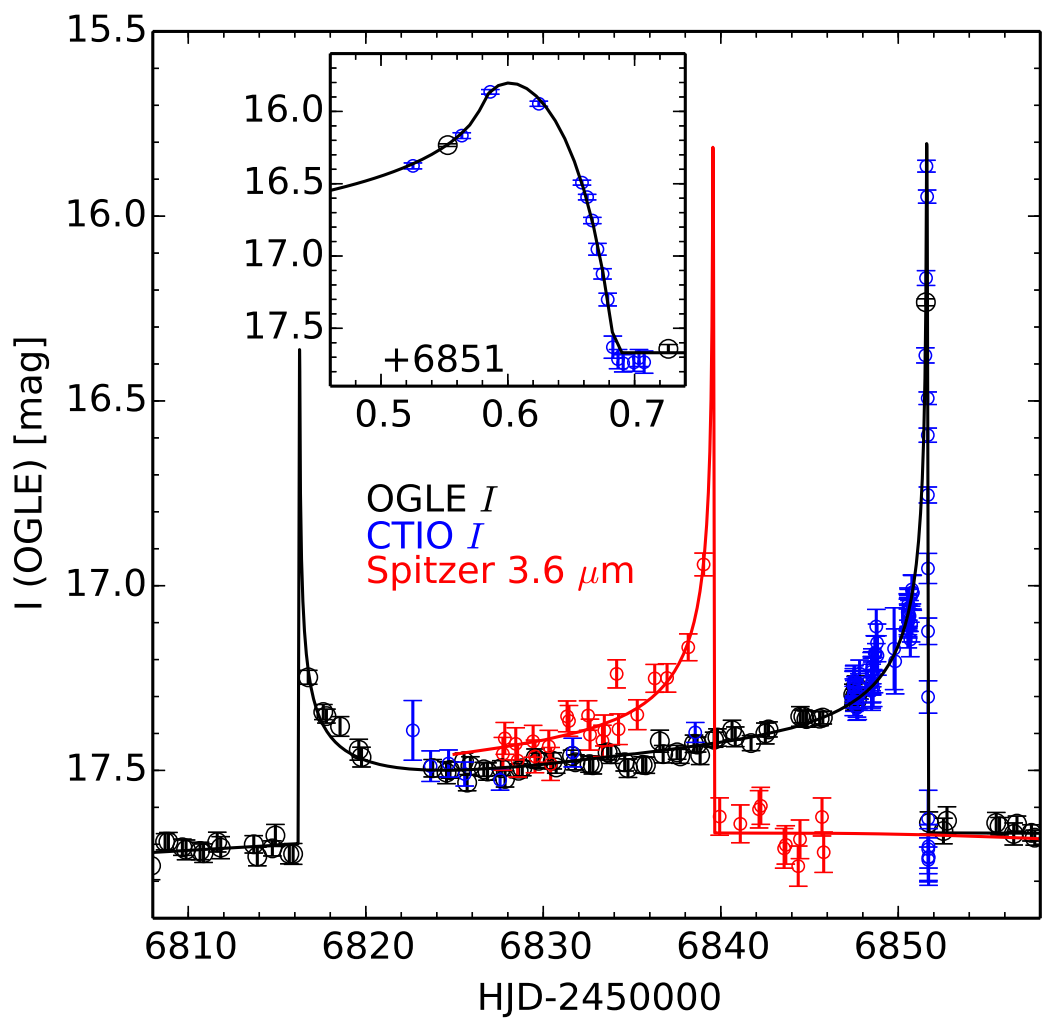}
\caption{\emph{Spitzer}'s parallax viewpoint of OGLE-2014-BLG-1050, while located $\sim1$ AU west of Earth. The very broad U-shaped trough in the ground-based light curve indicates a typical caustic-crossing binary-lens event. \emph{Spitzer} saw the same feature but $\sim 10$ days earlier, suggesting that the lens has a projected velocity of $\sim 200$ km s$^{-1}$ due East if \emph{Spitzer} traces the same trajectory as Earth. The inset shows the details of the caustic exit seen from Earth at $\sim6851$.
\label{fig:lc}}
\end{figure}

In total, we obtained 31 observations from \emph{Spitzer} from HJD$'$ = 6828 to HJD$'$ = 6846, which also happened to capture the caustic exit of the event. As shown in Figure~\ref{fig:lc}, the \emph{Spitzer} light curve shows very similar shape as the ground-based light curve, with the only significant difference being an offset of $\sim$10 days in time due to the microlens parallax effect, since Spitzer was displaced from the Earth by $\sim$1 AU.

\section{Light Curve Modeling} \label{sec:fitting}

The observed ground-based and space-based light curves of OGLE-2014-BLG-1050 are shown in Figure~\ref{fig:lc}. Only one major feature is noticed in the OGLE data: the prominent U-shaped trough, indicating a typical caustic-crossing binary-lens event. The \emph{Spitzer} data, on the other hand, only captured the caustic exit. The fact that the two lightcurves show similar shape and almost equal amplitude suggests that the trajectory of the source-lens relative motion as seen from \emph{Spitzer} should have similar impact parameter $u_0$ as that seen from Earth.

The four-fold degeneracy, coming from the fact that the satellite and Earth can pass on the same side, which we denote as $(+,+)$ and $(-,-)$ solutions with the two signs indicating the signs of $u_0$ as seen from Earth and \emph{Spitzer}, respectively \citep{Gould:2004}, or opposite sides of the lens, which we denote as $(+,-)$ and $(-,+)$ solutions accordingly, was well investigated in the case of single-lens microlensing \citep{Refsdal:1966,Gould:1994,GouldHorne:2013,Yee:2014,CalchiNovati:2014}.
\footnote{We note that these four solutions are defined in a different way in \citet{Yee:2014,CalchiNovati:2014}. The four solutions $(+,+)$, $(+,-)$, $(-,-)$ and $(-,+)$ by our definition correspond to $(-,+)$, $(+,+)$, $(-,-)$ and $(+,-)$ solutions by their definition, respectively.}
In the case of binary lenses in which the binary features are detected by both observatories, for example \citet{Udalski:2014}, this four-fold degeneracy normally collapses to the more general two-fold degeneracy, namely $u_{0,\pm}$ degeneracy, whose two solutions give very similar amplitudes of $\pi_{\rm E}$ and therefore $D_{\rm L}$ and $M_{\rm L}$ \citep{GouldHorne:2013}. However, due to the fact that \emph{Spitzer} covered only a small portion of the caustic-affected lightcurve, the persistence of such a four-fold degeneracy is suspected and confirmed in the current binary event.

Considering the characteristics of this event, we adopt a small variant of the standard modeling parameterization. The modeling of the light curves caused by binary lenses usually requires 11 system parameters, including seven basic parameters -- the time of the closest lens-source approach, $t_0$; the impact parameter normalized by the Einstein radius, $u_0$; the Einstein timescale, $t_{\rm E}$; the normalized source size, $\rho$; the normalized projected separation between the binary components, $s$; the mass ratio, $q$; and the angle from the binary-lens axis to the lens-source relative motion, $\alpha$ -- and four parameters to allow for higher-order effects, namely the microlens parallax, quantified by $\pi_{{\rm E},N}$ and $\pi_{{\rm E},E}$, and the binary-lens orbital motion, quantified by $d\alpha/dt$ and $ds/dt$. We refer the reader to \citet{Udalski:2014} for the sign definitions of parameters $u_0$, $\alpha$ and $d\alpha/dt$. 
In addition, each observatory requires two flux parameters $(f_{\rm S},f_{\rm B})$, so that the total flux is determined by
\[ f_{\rm tot}(t) = f_{\rm S} \cdot A(t) + f_{\rm B}\ ,\]
where $A(t)$ is the magnification of the source as a function of time $t$. These flux parameters are defined in a system in which $f=1$ corresponds to an 18 mag star.
For the current event, however, noticing that the time of caustic exit (as observed from Earth), $t_{\rm ce}$, is much better determined than $t_0$, we replace $t_0$ with $t_{\rm ce}$ as a free parameter, and then search for the $t_0$ that produces the given $t_{\rm ce}$ in order to compute the light curve \citep{Cassan:2008}. Parameters leading to an unbound binary system, quantitatively $\beta>1$, are not taken into account, in which \citep{Dong:2009}
\begin{equation}
\beta \equiv \left(\frac{E_{\rm kin}}{E_{\rm pot}}\right)_\perp = \frac{\kappa M_\odot ({\rm yr})^2}{8\pi^2} \frac{\pi_{\rm E} s^3 \gamma^2}{\theta_{\rm E} (\pi_{\rm E} + \pi_{\rm S}/\theta_{\rm E})^3};\quad  \bdv{\gamma} = (\gamma_\parallel,\gamma_\perp) \equiv \left( \frac{ds/dt}{s},\frac{d\alpha}{dt} \right)\ ;
\end{equation}
and the angular Einstein radius 
\[
    \theta_{\rm E} = \frac{\theta_{\star,\rm fid} (f_{\rm S}/f_{\rm S,fid})^{1/2}}{\rho}
\]
where $\theta_{\star,\rm fid}$ is the fiducial angular source size that is determined for a fiducial source flux $f_{\rm S,fid}$ (see Section~\ref{sec:cmd} for the final determination of $\theta_\star$).

The linear limb darkening coefficients we adopt, based on the source color, are $(\Gamma_I,\Gamma_{3.6})=(0.43,0.16)$, following the same procedures as in \citet{Udalski:2014}.

The Markov Chain Monte Carlo (MCMC) method is implemented to find the minimum and the likelihood distribution of parameters. As usual, the point source, quadrupole and hexadecapole \citep{Pejcha:2009,Gould:2008} approximations are used when the source is approaching, although still reasonably far (a few source radii) from, the caustics. For epochs that are near or on crossing caustics, we use contour integration, in which the limb darkening effect is accommodated by using 10 annuli \citep{GouldGaucherel:1997,Dominik:1998}.
\footnote{A more advanced and optimized version of the contour integration proposed by \citet{Bozza:2010} has been used in the modeling of real time data by V.B..}
However, this contour integration may fail at some particular points, in which case we use the more time-consuming inverse ray shooting \citep{Dong:2006}.

\begin{figure}
\centering
\plotone{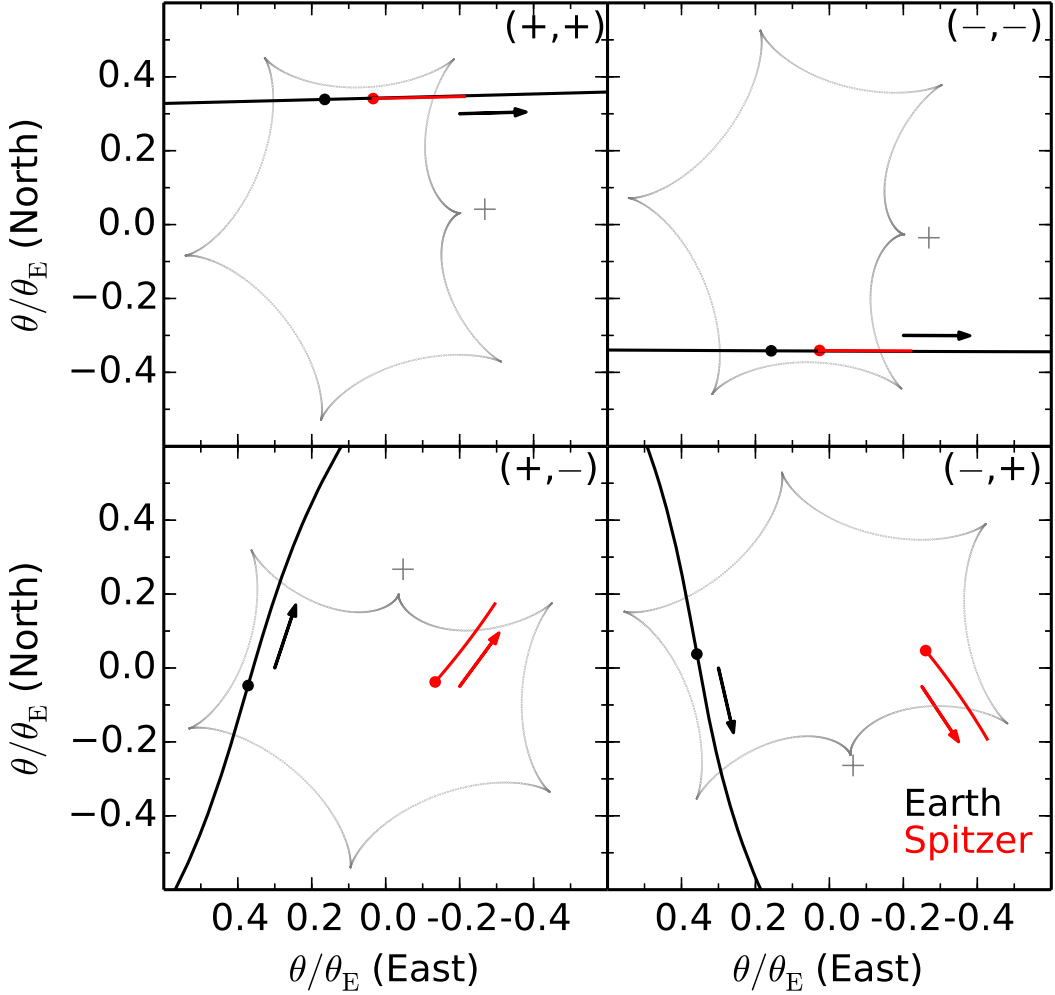}
\caption{Caustics (grey) and source-lens relative trajectories as seen from Earth (black line) and \emph{Spitzer} (red line); the red and black dots mark the source positions as seen from Earth and \emph{Spitzer} when \emph{Spitzer} started its observations on HJD$'\simeq 6828$ (the size of the dots here do not represent the source size); in each panel, the plus sign marks the position of the primary star, and the secondary is $\sim1.1$ Einstein units away along the direction of the axis of symmetry and on the opposite side of the caustic. In the favored solutions (upper panels), \emph{Spitzer} traces the same trajectory as Earth but $\sim 10$ days earlier, resulting in a similar light curve but displaced by $\sim 10$ days (see Figure~\ref{fig:lc}). However, similar lightcurves could also be induced if \emph{Spitzer} passed the lens on the opposite side but with similar impact parameters (lower panels). A much larger $\pi_{\rm E}$ is required in these cases, leading to the noticeable wiggles on the source trajectory as seen from Earth. 
\label{fig:cau}}
\end{figure}

We first fit only the ground-based data, i.e., the $I$ band data from OGLE and CTIO, in order to obtain a basic sense of this event. The best-fit parameters of the two solutions ($u_{0,\pm}$) are listed in Table~\ref{tab:ground-fitting}. Note in particular that the parallax parameters $\pi_{{\rm E},N}$ and $\pi_{{\rm E},E}$, are not significantly detected. Because the source lies very close to the ecliptic, the two solutions suffer from the ``ecliptic degeneracy'' \citep{Jiang:2004,Skowron:2011}. The \emph{Spitzer} data were then included, which yields much better constraints on the microlens parallax vector $\bdv{\pi}_{\rm E}=(\pi_{{\rm E},N},\pi_{{\rm E},E})$. To do so, one needs a careful initial setup of $\pi_{{\rm E},N}$ and $\pi_{{\rm E},E}$. 
That is, the \emph{Spitzer} caustic crossing clearly occurs between the data points on HJD$'$ = 6839.03 and HJD$'$ = 6839.94, so if the trial solutions do not have this property the $\chi^2$ minimization procedure will never arrive there.
We have
\begin{equation}
\bdv{\pi}_{\rm E} = \frac{\rm AU}{D_\perp} \left( \frac{\Delta t_0}{t_{\rm E}}, \Delta u_0 \right)\ ,
\end{equation}
in which $\Delta t_0 = t_{0,\rm sat}-t_{0,\oplus}$, $\Delta u_0 = u_{0,\rm sat}-u_{0,\oplus}$, and $\bdv{D}_\perp$ is the projected separation vector of the Earth and satellite \citep{Udalski:2014}. For solutions in which Earth and \emph{Spitzer} pass the lens on the same side, the similarity between the ground-based and \emph{Spitzer} lightcurves suggests that $\Delta u_0 \approx 0$, which leads to $\pi_{{\rm E},N} \approx 0$. Then, since $D_\perp \approx 1$ AU, $\Delta t_0 \sim 10$ days, and $t_{\rm E} \approx 80$ days, one obtains $\pi_{{\rm E},E} \approx 0.13$. For the other two solutions, the situation is less straightforward because of the difficulty in determining the direction of $\bdv{D}_\perp$ by simple inspection. In principle, one can still estimate $\bdv{\pi}_{\rm E}$ by working out the geometry more carefully, but a much easier approach is to conduct a grid search on $(\pi_{{\rm E},N},\pi_{{\rm E},E})$, with all other parameters initially set at the best-fit to ground-based data, so as to find a reasonably good starting points for MCMC sampling. The best-fit parameters for all four solutions of the fit to the combined data sets are listed in Table~\ref{tab:spitzer-fitting}, and the caustic structures and the lens-source relative motion as seen from Earth and \emph{Spitzer} are shown in Figure~\ref{fig:cau}.

As expected, the inclusion of \emph{Spitzer} data reduces the error bars on $\pi_{{\rm E},N}$ and $\pi_{{\rm E},E}$ significantly. 
The four-fold degeneracy does persist in this binary-lens event, and the four solutions have nearly equal $\chi^2$, mostly due to the coincidence that the source-lens relative motion is almost parallel to the binary-lens direction, i.e., $\alpha \sim 180^\circ$, and also the fact that \emph{Spitzer} data do not have a long enough time baseline. However, this four-fold degeneracy is effectively broken by other considerations, as we will discuss in Section~\ref{sec:physical}.

The inclusion of the lens orbital motion effect is important in order to uncover the true physical parameters of the lens system \citep{Park:2013}. For OGLE-2014-BLG-1050, the orbital motion effect is only marginally detected ($\Delta \chi^2 \approx 5$), but it enlarges the uncertainty on parameters such as $q$ and $u_0$ by a factor of two to three. 

\begin{figure}
\plotone{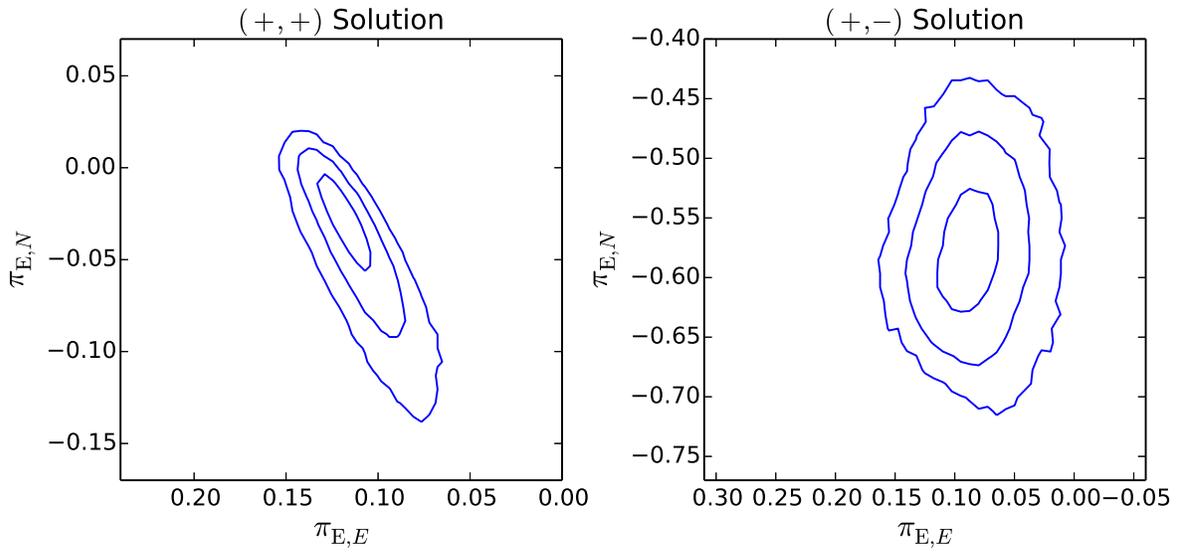}
\caption{The 2D posteriors between $\pi_{{\rm E},N}$ and $\pi_{{\rm E},E}$ for the $(+,+)$ and $(+,-)$ solutions. The contours (representing 1-$\sigma$, 2-$\sigma$ and 3-$\sigma$ limits) enclose probabilities of $39\%$, $86\%$ and $99\%$, respectively. Compared to the $(+,-)$ solution, the $(+,+)$ solution has much more asymmetric contours in the $(\pi_{{\rm E},N},\pi_{{\rm E},E})$ plane.
\label{fig:posteriors}}
\end{figure}

It is interesting to compare this binary-lens event with the only planetary event so far found with the same method, OGLE-2014-BLG-0124 \citep{Udalski:2014}. One noticeable difference between these two events is the uncertainty in $\pi_{\rm E}$. In the present binary-lens event, the uncertainty in $\pi_{\rm E}$ ($10-20\%$) after the inclusion of \emph{Spitzer} data is considerably larger than the one for the planetary event OGLE-2014-BLG-0124 ($\sim 2.5\%$). The reason is that the \emph{Spitzer} light curve of OGLE-2014-BLG-1050 has fewer features, which itself derives from two facts. The first is that due to the Sun-angle limitation, \emph{Spitzer}'s total time baseline is not long enough to cover the caustic entrance as was the case for the planetary event. The second is that the current program limits the observation cadence to no more than once per day, and thus we were not able to capture the details of the caustic exit.
Another consequence of this single-feature \emph{Spitzer} light curve is the asymmetric posterior distributions of $\pi_{{\rm E},N}$ and $\pi_{{\rm E},E}$ in the $(+,+)$ and $(-,-)$ solutions. See Figure~\ref{fig:posteriors} for the 2D posteriors of $(\pi_{{\rm E},N},\pi_{{\rm E},E})$ from the $(+,+)$ solution with respect to that from the $(+,-)$ solution. This is because \emph{Spitzer}'s view of the source-lens relative trajectory in the $(+,+)$ [$(-,-)$] solution can go further in the direction of South (North) than North (South). However, this restriction is much weaker for the other two solutions (see Figure~\ref{fig:cau}). We show in the Appendix the triangle diagrams of the fitting parameters and derived physical parameters for the two solutions $(+,\pm)$.

\section{Color Magnitude Diagram} \label{sec:cmd}

We use a variant of the standard procedure to determine the angular size of the source star \citep{Yoo:2004}. The standard procedure usually requires magnified images of the event taken in both $V$ and $I$ bands. In the case of OGLE-2014-BLG-1050, $\mu$FUN did not take any $V$-band images when the source was substantially magnified. 
Instead, we determine the $V-I$ color from the $I-H$ color following the procedures introduced by \citet{Yee:2013}, by taking advantage of the fact that the SMARTS camera takes $H$-band images simultaneously with the $I$-band images. The instrumental $I-H$ color of the source is first determined by linear regression of $H$ on the $I$ flux at various magnifications during the event. The $V-I$ color is then found by using the color-color relation derived from nearby field stars. Since nearby field stars are mostly redder than the source, we have to extrapolate the color-color relation blueward to the source position, which introduces a 0.05 mag error in the source $V-I$ color. 
The instrumental $I$-band baseline flux of the source is determined from the modeling ($f_{\rm S,CTIO} = 0.1612 \pm 0.0016$). Therefore, the instrumental color and magnitude of the source are determined to be $(V-I,I)_{\rm S,CTIO} = (-0.73,19.98)$. We then compare the source with the centroid of the red clump and find an offset of $\Delta (V-I,I)_{\rm S} = (-0.31,3.77)$.
With the intrinsic centroid of the red clump \citep{Bensby:2013,Nataf:2013} and a distance modulus of 14.44 \citep{Nataf:2013}, we determine the dereddened source color and magnitude to be $(V-I,I)_{\rm S,0} = (0.75,18.09)$, and $M_I=3.65$, making the source a turn-off star.

To determine the source angular size, we then convert from $V-I$ to $V-K$ using the empirical color-color relations of \citet{BessellBrett:1988}
\footnote{In principle, one can directly convert the $I-H$ color to the $V-K$ color. This is not adopted in the present work since the centroid of the red clump is not determined in the $(I,I-H)$ plane so well as in the $(I,V-I)$ plane, and the determination of the red clump centroid in the $(I,I-H)$ plane is not within the scope of the current work.}
, apply the color/surface-brightness relation of \citet{Kervella:2004}, and finally find
\[ \theta_\star = 0.80 \pm 0.09\ \mu{\rm as}\ .\]
The error comes from the extrapolation of the color-color relation to the source regime ($\sim0.05$ mag), the determination of the red clump centroid using nearby field stars ($\sim0.05$ mag) and the derivation of the intrinsic source color \citep[0.05 mag,][]{Bensby:2013}.

The lightcurve modeling shows that the source is severely blended. Hence, we also determine the color and magnitude of the ``blend'' (blending light) so as to test whether it can be explained by the lens system. From OGLE IV field star photometry, the ``baseline object'' (source plus blend) is found to be $(V-I,I)_{\rm base,OGLE} = (2.09,17.81)$, and the centroid of the red clump $(V-I,I)_{\rm RC,OGLE} = (2.75,16.27)$. The correction of non-standard $V$ band in OGLE IV is then applied
\[ \Delta (V-I)_{\rm JC} = 0.92 \cdot \Delta (V-I)_{\rm OGLE} ,\]
in which `JC' represents the standard Johnson-Cousins system. With the dereddened red clump at $(V-I,I)_{\rm RC,JC} = (1.06,14.32)$ and assuming the same extinction law $R_I \equiv A_I/E(V-I)=1.23$ \citep[the distance modulus and $R_V$ are taken from][at the event position]{Nataf:2013}, the total flux baseline and the source are determined to be $(V-I,I)_{\rm base,JC} = (2.04,17.81)$ and $(V-I,I)_{\rm S,JC} = (2.44,20.08)$ respectively. We then find the color and magnitude of the blend to be $(V-I,I)_{\rm B,JC} = (2.01,17.95)$.

We show in Figure~\ref{fig:cmd} the CMD that is used to characterize the source and blend, after correction to the standard Johnson-Cousins system, together with the color and magnitude of an example lens system from the preferred physical solution (see Section~\ref{sec:physical}).

\section{Physical Parameters} \label{sec:physical}

The physical parameters derived from our modeling are given in Table~\ref{tab:spitzer-orbit-physical}, in which $\bdv{\tilde{v}}_{\rm hel}=(\tilde{v}_{{\rm hel},N},\tilde{v}_{{\rm hel},E})$ is the projected velocity of the lens relative to the source in the heliocentric frame, coming from
\[ \bdv{\tilde{v}}_{\rm hel} = \frac{\bdv{\pi}_{\rm E} \rm AU}{\pi_{\rm E}^2 t_{\rm E}} + \bdv{v}_{\oplus,\perp} \]
with $\bdv{v}_{\oplus,\perp}\approx(0.7,28.3)$ km s$^{-1}$ being the velocity of Earth projected on the sky at the peak of the event. The four-fold degeneracy basically collapses to two physical solutions, since the two solutions $(+,+)$ and $(-,-)$ [collectively $(\pm,\pm)$] have the same amplitude of parallax $\pi_{\rm E}$ [as do the other two solutions $(\pm,\mp)$] and therefore lead to similar lens system properties.

Our results show that the microlensing event OGLE-2014-BLG-1050 was produced by a binary system consisting of $0.9 M_\odot$ and $0.35 M_\odot$ stars separated by 5 AU at a distance 3.5 kpc from the Sun (high-mass binary solution), or consisting of $0.2 M_\odot$ and $0.07 M_\odot$ stars separated by 1.6 AU at a distance 1.1 kpc from the Sun (low-mass binary solution). 

The Rich argument, originally suggested by James Rich (circa 1997, private communication) and elaborated in \citet{CalchiNovati:2014}, 
argues that if the two components of $\bdv{\pi}_{\rm E}D_\perp/$AU (namely $\Delta t_0/t_{\rm E}$ and $\Delta u_0$) are small and of the same order, then the $(\pm,\pm)$ solutions are strongly favored over the $(\pm,\mp)$ solutions because the latter require fine tuning.  However, because this argument rests on the axial symmetry of the point-lens geometry, it cannot be applied to binary lenses in a straightforward manner.

\begin{figure}
\centering
\plotone{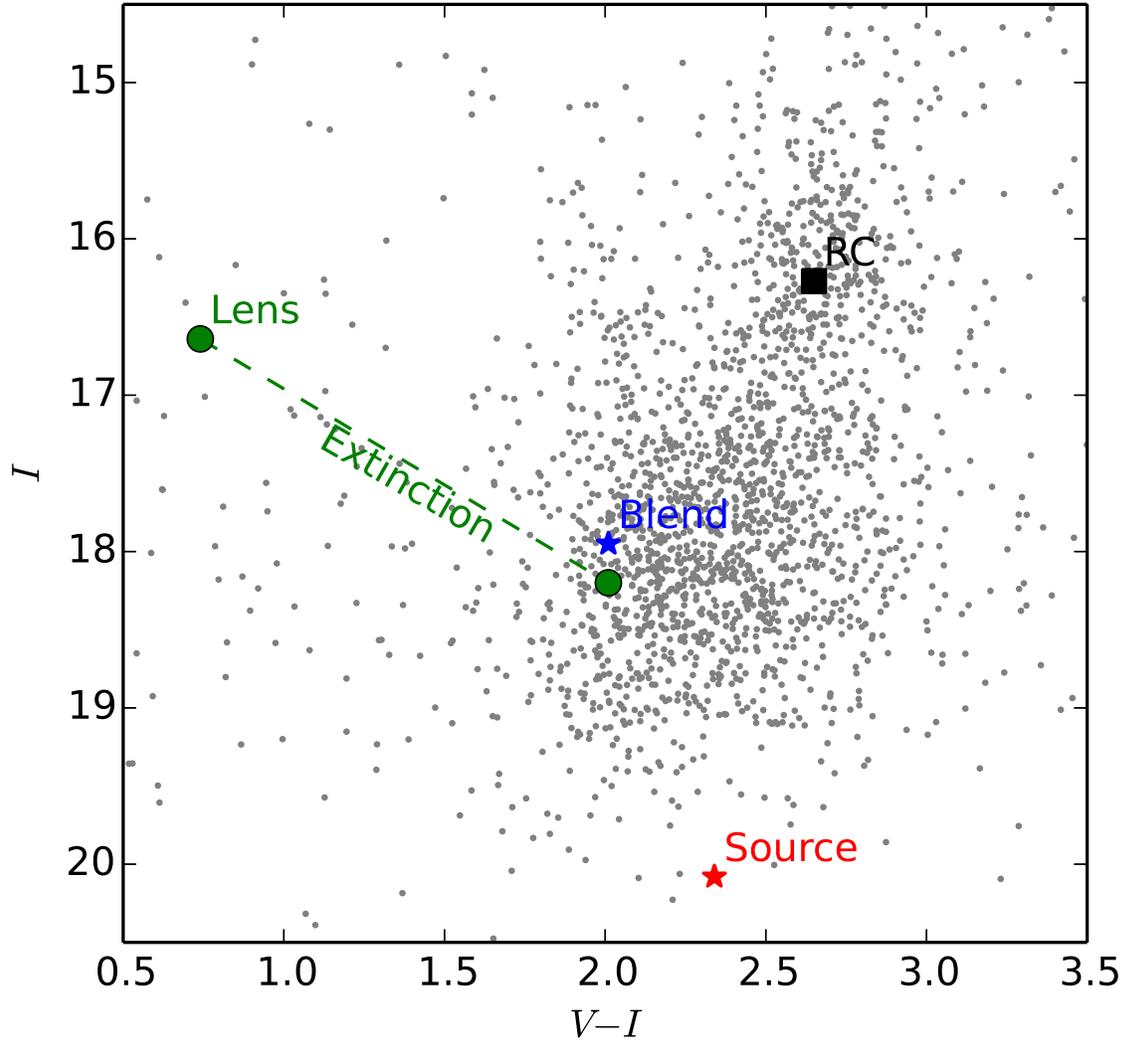}
\caption{The color-magnitude diagram of stars in a $3.8'\times3.8'$ square centered on the source star. The positions of the centroid of the red clump (`RC'), the source star, the blended light, and an example lens system taken from the high-mass binary solution before and after the extinction are marked. 
\label{fig:cmd}}
\end{figure}

As seen in Table~\ref{tab:spitzer-orbit-fitting}, the two different physical solutions have nearly equal $\chi^2$. However, the color and magnitude of the blend determined in Section~\ref{sec:cmd} strongly support the high-mass binary solution. 
We demonstrate this point qualitatively by taking a Sun-like primary with a $0.4M_\odot$ secondary at 3.2 kpc, all of which are well within the 1-$\sigma$ error bar of the high-mass binary solution. The primary would have a color $V-I=0.70$ \citep{Ramirez:2012} and magnitude $M_I=4.15$. For the secondary, we take $V-I=2.5$ and $M_I=7.6$. Together with the assumed distance 3.2 kpc, these give the combined dereddened color and magnitude $(V-I,I)_{\rm L,0} = (0.74,16.64)$. 
With the same $R_V$ as we used in Section~\ref{sec:cmd} assumed, in order to match the color of the blend, the lens system should suffer an extinction with $E(V-I)=1.27$ and therefore $A_I=R_I \cdot E(V-I)=1.56$. This leads to an apparent $I$ band magnitude of $18.20$ for the lens system. Although it is still 0.25 mag fainter than the total blend, we emphasize that this is just a qualitative demonstration of the consistency between the lightcurve modeling and photometry. In fact, many factors could be used to explain this discrepancy, such as metallicity, stellar evolution stage, or a more massive and closer binary system
\footnote{Note that a more massive and closer binary system is possible since the error bars on the derived mass and distance are relatively large. For example, raising $\theta_{\rm E}$ by $5\%$ increases the mass and relative parallax both by $5\%$, leading to a roughly $23\%$ brightness increment.}
. We show in Figure~\ref{fig:cmd} this example track of the lens system on the CMD.

In principle, the blend can be a nearby field star other than the lens. However, this scenario is extremely unlikely for this event. Using the OGLE data taken both before and during the event, we are able to independently determine the light centroids of the source and the ``baseline object''. These give an offset of $42$ mas, meaning that the source of the excess flux must lie within this limit (to be conservative, we use $100$ mas for the following estimate). Note that this difference (0.15 pixels) is consistent with being coincident because there is a star of comparable brightness to the blend that lies just 0.9$''$ away.
With the measured color and magnitude of the excess flux, $(V-I,I)_{\rm B,JC} = (2.01,17.95)$, we search for stars whose brightness is within $I_{\rm B,JC} \pm 0.5$ mag and color within $(V-I)_{\rm B,JC}\pm0.3$, and find 344 such stars from the CMD shown in Figure~\ref{fig:cmd}. Recalling that this CMD is made using all stars within a $3.8' \times 3.8'$ square centered on the microlensing event, this indicates that the probability for such a star to lie within 100 mas from the source is only $0.02\%$.

A complementary argument that also supports the high-mass binary solution is the relative lensing probability. 
The event rate $\Gamma=n \sigma v$ as a function of the independent physical variables $(M_{\rm L}, D_{\rm L}, \bdv{\mu}_{\rm rel})$ for single-lens microlensing event is \citep{Batista:2011}
\begin{equation} \label{eq:rate}
p_{\rm single} \propto \frac{d^4 \Gamma}{d D_{\rm L} d\log{M_{\rm L}} d^2\bdv{\mu}_{\rm rel}} = \nu(R,z) (2R_{\rm E}) v_{\rm rel} f(\bdv{\mu}_{\rm rel}) g(M_{\rm L}) 
\propto D_{\rm L}^2 \nu(R,z) f(\bdv{\mu}_{\rm rel}) M_{\rm L}^{-\alpha}  \ ,
\end{equation}
where $\nu(R,z)$ is the number density of stars at position $(R,z)$ relative to the Galactic center, $R_{\rm E} \equiv D_{\rm L} \theta_{\rm E}$ is the Einstein radius at the lens plane, $v_{\rm rel} \equiv D_{\rm L} \mu_{\rm rel}$ is the lens-source relative velocity, $f(\bdv{\mu}_{\rm rel})$ is the two-dimensional probability function for a given source-lens relative proper motion $\bdv{\mu}_{\rm rel}$, and $g(M_{\rm L})\equiv M_{\rm L}^{-\alpha}$ is the Galactic stellar mass function in equal bins of $\log{M}$. We take $\nu(R,z)$ in the form of
\[ \nu(R,z) \propto e^{-R/R_\star} e^{-|z|/H}\ , \]
and choose $R_\star \simeq 3$ kpc and $H \simeq 250$ pc; given the Galactic coordinates of this event, $\nu(R,z)$ yields a factor of 1.4 favoring the high-mass binary solution. We choose $\alpha=1$ for the power-law index of the mass function. For the relative proper motion, $\mu_{\rm rel}$ has equal amplitude for each solution, but its direction results in a factor of $\sim 3$ in $f(\bdv{\mu})$, favoring the low-mass binary solution \citep{Yee:2014,CalchiNovati:2014}.
For binary-lens event, there should be three factors in addition to those in Equation~(\ref{eq:rate}). The multiplicity frequency, $f_{\rm m}$, for G-type (primary of the high-mass solution) and M-type (the primary of low-mass solution) dwarfs, is different by a factor of $\sim 2$ \citep{DucheneKraus:2013}. The mass ratio $q$ results in a factor of unity, since both solutions have the same $q$. The semi-major axis $a$ also leads to a factor of unity, given an \"Opik law in $\log{a}$ and that $a \simeq \sqrt{3/2} a_\perp$ \citep{Zhu:2014}. With all the above factors as well as the distance factor $D_{\rm L}^2$ considered, this argument leads to a conclusion that the high-mass distant binary is more likely to be microlensed than the low-mass close binary by a factor of $\sim 2$.

The blend origin strongly prefers the high-mass binary solution, and this solution is also preferred by the relative lensing probability. Nevertheless, we suggest that future direct imaging of the event with adaptive optics will help to reach a definitive conclusion about the nature of the lens system \citep{Batista:2014}.

\section{Discussion} \label{sec:discussion}

After 50 years of dreaming, the concept originally proposed by \citet{Refsdal:1966} to probe the mass function of Galactic astronomical objects without biases toward brightness is finally underway. With combined observations from \emph{Spitzer} and ground, we have shown that the mass and distance of the microlensing planetary events can be well constrained \citep[see][]{Udalski:2014}, and demonstrated the potential of using microlens parallax to probe the Galactic distribution of planets \citep[see][]{CalchiNovati:2014}.

In the present work, we report on a binary-lens event OGLE-2014-BLG-1050 observed in our program to demonstrate the power of using space-based microlens parallax to measure the mass and distance of binaries. Binary-lens events attract special attention because the finite source effect is often detected during caustic crossings and thus leads to well constrained Einstein ring radius $\theta_{\rm E}$.

Unlike the planetary event OGLE-2014-BLG-0124, in which the microlens parallax parameter $\pi_{\rm E}$ is well determined \citep[$2.5\%$ uncertainty,][]{Udalski:2014},
$\pi_{\rm E}$ is only constrained to within $\sim20\%$ in the present binary-lens event because of its single-feature \emph{Spitzer} light curve. 
The caustic entrance was not captured by \emph{Spitzer} for two reasons. First, by the time of OGLE's alert, the entrance had already occurred as seen by \emph{Spitzer}. Second, given \emph{Spitzer}'s Sun-angle limitation it would have been impossible to extend the total time baseline to capture the caustic entrance even with an earlier alert. However, the uncertainty in $\pi_{\rm E}$ could still have been reduced significantly if more \emph{Spitzer} observations had been obtained during the caustic exit. Nevertheless, we emphasize that with \emph{Spitzer} data the measurement of $\pi_{\rm E}$ is quite secure. By contrast, $\pi_{\rm E}$ is not significantly detected if only ground-based data are used.

Another interesting characteristic of OGLE-2014-BLG-1050 is that the four-fold degeneracy, which usually appears in single-lens events but has not been investigated in the binary-lens case, is unexpectedly present. This is mostly due to the coincidence that the source-lens relative motion is close to parallel to the binary-lens direction but also to the fact that \emph{Spitzer} data did not capture the caustic entrance. 
\footnote{We remind that this four-fold degeneracy will disappear for programs with long enough time baseline.}
The resulting four degenerate solutions are almost equal in $\chi^2$ and lead to two very different physical solutions for the mass and distance of the lens system: a binary system consisting of $0.9M_\odot$ and $0.35M_\odot$ stars separated by 5 AU at 3.5 kpc (the high-mass binary solution), or a binary consisting of $0.2M_\odot$ and $0.07M_\odot$ stars separated by 1.6 AU at 1.1 kpc (low-mass binary solution).

However, this degeneracy is effectively broken when two other factors are considered. The color and magnitude of the blend, after the source contribution is subtracted from the total baseline, can be well explained by the high-mass binary solution within its 1-$\sigma$ error bars. By contrast, the chance that a random field star is responsible for this blend is only 0.02\%. The lensing probability estimate also favors the high-mass binary solution by a factor of two.

The microlensing event OGLE-2014-BLG-1050 demonstrates the power of microlens parallax in measuring mass and distance of binaries. Future space-based programs with \emph{Spitzer} \citep{Gould:2014} and future missions such as \emph{Kepler} \citep[K2,][]{GouldHorne:2013}, Euclid \citep{Penny:2013} and the Wild Field InfraRed Survey Telescope \citep{Spergel:2013} can help draw a full picture of the Galactic distribution of binary systems, from brown dwarf binaries to binaries involving black holes.

\acknowledgments

Work by WZ, AG and BSG were supported by NSF grant AST 1103471.
Work by JCY, AG and SC was supported by JPL grant 1500811.
AG, BSG, and RWP were supported by NASA grant NNX12AB99G. 
Work by CH was supported by the Creative Research Initiative Program (2009-0081561) of the National Research Foundation of Korea.
Work by JCY was performed under contract with the California Institute of Technology (Caltech)/Jet Propulsion Laboratory (JPL) funded by NASA through the Sagan Fellowship Program executed by the NASA Exoplanet Science Institute.
Work by CAB was carried out in part at the Jet Propulsion Laboratory (JPL), California Institute of Technology, under a contract with the National Aeronautics and Space Administration.
The OGLE project has received funding from the European Research Council under the European Community’s Seventh Framework Programme (FP7/2007-2013) / ERC grant agreement no. 246678 to AU.
This work is based in part on observations made with the Spitzer Space Telescope, which is operated by the Jet Propulsion Laboratory, California Institute of Technology under a contract with NASA.


\clearpage
\begin{deluxetable}{lcc}
    \centering
    \tablecaption{Best-fit parameters for ground-based only fit.
    \label{tab:ground-fitting}}
    \tablehead{Parameters & $u_0>0$ (`$+$' solution) & $u_0<0$ (`$-$' solution)}
    \startdata
    $\chi^2$/dof & 457.6/506 & 458.4/506 \\
$t_{\rm ce}$ (HJD$^\prime-$6851) & 0.6318 $\pm$ 0.0005 & 0.6317 $\pm$ 0.0005 \\
$u_0$ & 0.357 $\pm$ 0.014 & 0.354 $\pm$ 0.017 \\
$t_{\rm E}$ (days) & 79.2 $\pm$ 6.4 & 78.0 $\pm$ 4.3 \\
$\rho$ ($10^{-4}$) & 6.09 $\pm$ 0.64 & 6.23 $\pm$ 0.50 \\
$\pi_{{\rm E},N}$ & -0.45 $\pm$ 0.40 & 0.36 $\pm$ 0.45 \\
$\pi_{{\rm E},E}$ & 0.31 $\pm$ 0.24 & 0.40 $\pm$ 0.26 \\
$\alpha$ (deg) & 180.3 $\pm$ 2.4 & 179.8 $\pm$ 2.3 \\
$ds/dt$ (yr$^{-1}$) & -- & -- \\
$d\alpha/dt$ (yr$^{-1}$) & --  & -- \\
$s$ & 1.108 $\pm$ 0.028 & 1.115 $\pm$ 0.023 \\
$q$ & 0.357 $\pm$ 0.052 & 0.371 $\pm$ 0.031 \\

    \enddata
\end{deluxetable}

\clearpage
\begin{deluxetable}{lcccc}
    \centering
    \tablecaption{Best-fit parameters for the fit to the combined datasets (no orbital motion).
    \label{tab:spitzer-fitting}}
    \tablehead{Parameters & $(+,+)$ & $(-,-)$ & $(+,-)$ & $(-,+)$}
    \startdata
    $\chi^2$/dof & 489.5/506 & 489.6/506 & 489.5/506 & 490.4/506 \\
$t_{\rm ce}$ (HJD$^\prime-$6851) & 0.63174 $\pm$ 0.00050 & 0.63173 $\pm$ 0.00050 & 0.63157 $\pm$ 0.00049 & 0.6317 $\pm$ 0.00049 \\
$u_0$ & 0.352 $\pm$ 0.015 & 0.353 $\pm$ 0.015 & 0.358 $\pm$ 0.014 & 0.361 $\pm$ 0.016 \\
$t_{\rm E}$ (days) & 76.7 $\pm$ 3.9 & 76.9 $\pm$ 4.0 & 77.5 $\pm$ 3.6 & 77.9 $\pm$ 4.0 \\
$\rho$ ($10^{-4}$) & 6.25 $\pm$ 0.47 & 6.22 $\pm$ 0.47 & 6.15 $\pm$ 0.45 & 6.07 $\pm$ 0.47 \\
$\pi_{{\rm E},N}$ & -0.037 $\pm$ 0.031 & 0.029 $\pm$ 0.032 & -0.577 $\pm$ 0.056 & 0.571 $\pm$ 0.057 \\
$\pi_{{\rm E},E}$ & 0.115 $\pm$ 0.016 & 0.117 $\pm$ 0.016 & 0.089 $\pm$ 0.029 & 0.098 $\pm$ 0.029 \\
$\alpha$ (deg) & 180.3 $\pm$ 1.6 & 180.6 $\pm$ 1.6 & 178.9 $\pm$ 1.8 & 178.5 $\pm$ 1.7 \\
$ds/dt$ (yr$^{-1}$) & -- & -- & -- & --  \\
$ds/dt$ (yr$^{-1}$) & -- & -- & -- & --  \\
$s$ & 1.123 $\pm$ 0.020 & 1.123 $\pm$ 0.020 & 1.116 $\pm$ 0.019 & 1.114 $\pm$ 0.021 \\
$q$ & 0.382 $\pm$ 0.029 & 0.380 $\pm$ 0.029 & 0.383 $\pm$ 0.028 & 0.383 $\pm$ 0.028 \\

    \enddata
\end{deluxetable}

\begin{deluxetable}{lcccc}
    \centering
    \tablecaption{Best-fit parameters for the fit to the combined datasets (with orbital motion).
    \label{tab:spitzer-orbit-fitting}}
    \tablehead{Parameters & $(+,+)$ & $(-,-)$ & $(+,-)$ & $(-,+)$}
    \startdata
    $\chi^2$/dof & 484.7/506 & 485.4/506 & 483.6/506 & 483.5/506 \\
$t_0$ (HJD$^\prime-$6851) & 0.63170 $\pm$ 0.00050 & 0.63176 $\pm$ 0.00048 & 0.63166 $\pm$ 0.00050 & 0.63181 $\pm$ 0.00051 \\
$u_0$ & 0.352 $\pm$ 0.036 & 0.347 $\pm$ 0.030 & 0.343 $\pm$ 0.034 & 0.331 $\pm$ 0.033 \\
$t_{\rm E}$ (days) & 73.3 $\pm$ 4.2 & 73.3 $\pm$ 4.2 & 70.0 $\pm$ 3.8 & 70.3 $\pm$ 4.3 \\
$\rho$ ($10^{-4}$) & 6.05 $\pm$ 0.64 & 6.11 $\pm$ 0.59 & 6.08 $\pm$ 0.65 & 6.06 $\pm$ 0.58 \\
$\pi_{{\rm E},N}$ & -0.046 $\pm$ 0.051 & 0.036 $\pm$ 0.036 & -0.571 $\pm$ 0.063 & 0.556 $\pm$ 0.066 \\
$\pi_{{\rm E},E}$ & 0.111 $\pm$ 0.025 & 0.115 $\pm$ 0.020 & 0.081 $\pm$ 0.035 & 0.098 $\pm$ 0.031 \\
$\alpha$ (deg) & 178.8 $\pm$ 2.1 & 178.8 $\pm$ 2.0 & 176.9 $\pm$ 2.3 & 176.5 $\pm$ 2.1 \\
$ds/dt$ (yr$^{-1}$) & -0.10 $\pm$ 0.40 & -0.19 $\pm$ 0.39 & -0.37 $\pm$ 0.49 & -0.60 $\pm$ 0.52 \\
$d\alpha/dt$ (yr$^{-1}$) & 0.57 $\pm$ 0.27 & -0.53 $\pm$ 0.27 & 0.90 $\pm$ 0.53 & -0.77 $\pm$ 0.52 \\
$s$ & 1.113 $\pm$ 0.028 & 1.113 $\pm$ 0.025 & 1.105 $\pm$ 0.027 & 1.097 $\pm$ 0.031 \\
$q$ & 0.389 $\pm$ 0.094 & 0.367 $\pm$ 0.078 & 0.354 $\pm$ 0.093 & 0.314 $\pm$ 0.090 \\

    \enddata
\end{deluxetable}


\clearpage
\begin{deluxetable}{lcccc}
    \centering
    \tablecaption{Physical parameters (ground-based + \emph{Spitzer}, with orbital motion).
    \label{tab:spitzer-orbit-physical}}
    \tablehead{Parameters & $(+,+)$ & $(-,-)$ & $(+,-)$ & $(-,+)$}
    \startdata
    $M_{\rm primary}$ ($M_\odot$) & 0.91 $\pm$ 0.17 & 0.94 $\pm$ 0.17 & 0.21 $\pm$ 0.03 & 0.22 $\pm$ 0.04 \\
$M_{\rm secondary}$ ($M_\odot$) & 0.355 $\pm$ 0.079 & 0.342 $\pm$ 0.092 & 0.073 $\pm$ 0.016 & 0.067 $\pm$ 0.013 \\
$D_{\rm L}$ (kpc) & 3.46 $\pm$ 0.37 & 3.47 $\pm$ 0.19 & 1.12 $\pm$ 0.17 & 1.15 $\pm$ 0.17 \\
$a_\perp$ (AU) & 5.02 $\pm$ 0.43 & 5.04 $\pm$ 0.37 & 1.64 $\pm$ 0.14 & 1.66 $\pm$ 0.14 \\
$\theta_{\rm E}$ (mas) & 1.34 $\pm$ 0.16 & 1.32 $\pm$ 0.14 & 1.33 $\pm$ 0.15 & 1.32 $\pm$ 0.13 \\
$\tilde{v}_{{\rm hel},N}$ (km/s) & -75 $\pm$ 46 & 58 $\pm$ 46 & -42 $\pm$ 4 & 42 $\pm$ 5 \\
$\tilde{v}_{{\rm hel},E}$ (km/s) & 205 $\pm$ 47 & 208 $\pm$ 32 & 35 $\pm$ 2 & 37 $\pm$ 2 \\

    \enddata
\end{deluxetable}

\clearpage

\clearpage

\clearpage

\clearpage

\clearpage

\appendix
\section{Appendix material}

\begin{figure}
\centering
\plotone{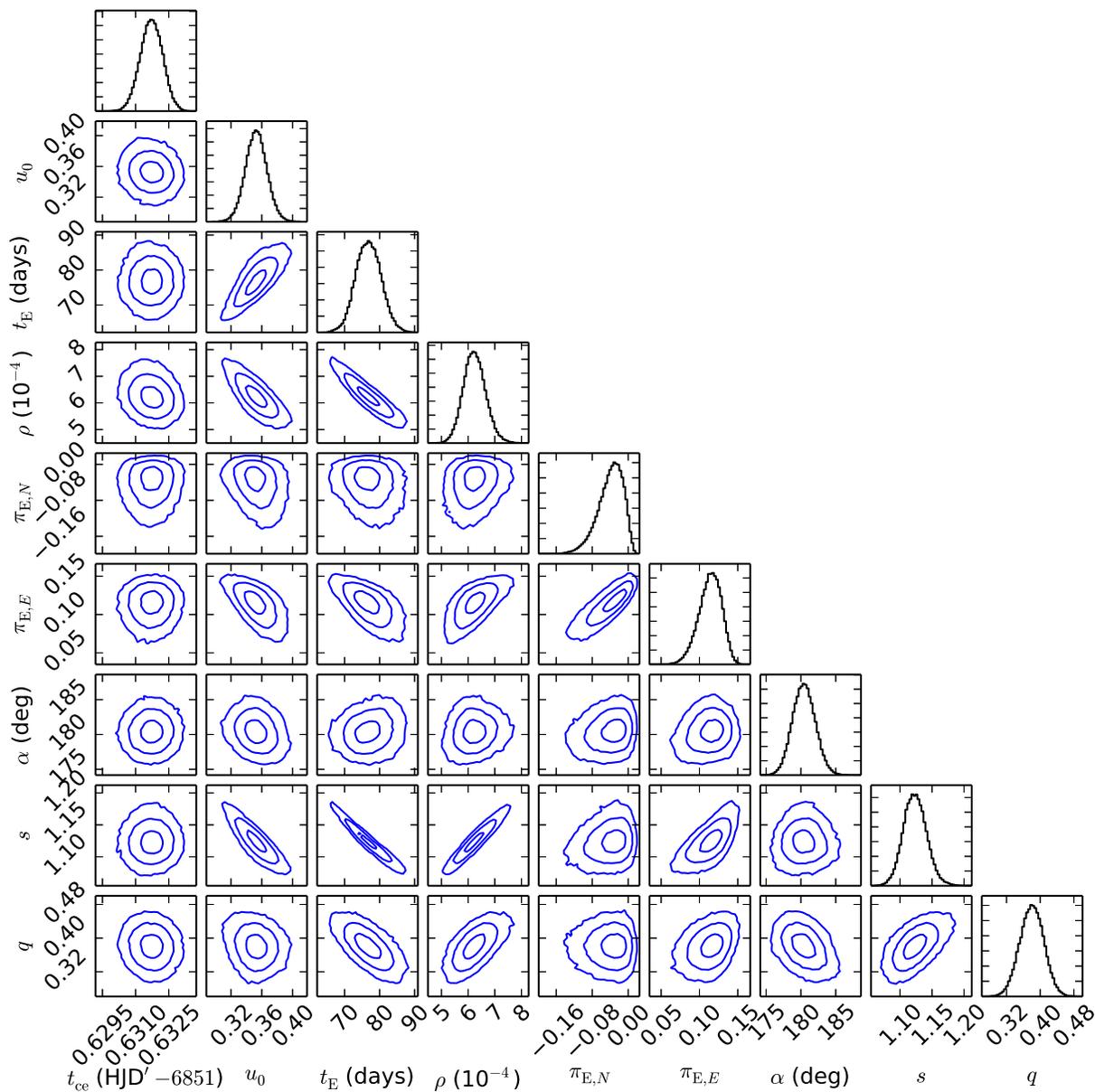}
\caption{The full triangle diagram of the fitting parameters in the $(+,+)$ solution without orbital motion.
The contours (representing 1-$\sigma$, 2-$\sigma$ and 3-$\sigma$ limits) enclose probabilities of $39\%$, $86\%$ and $99\%$, respectively.
\label{fig:full-plus}}
\end{figure}

\clearpage
\begin{figure}
\centering
\plotone{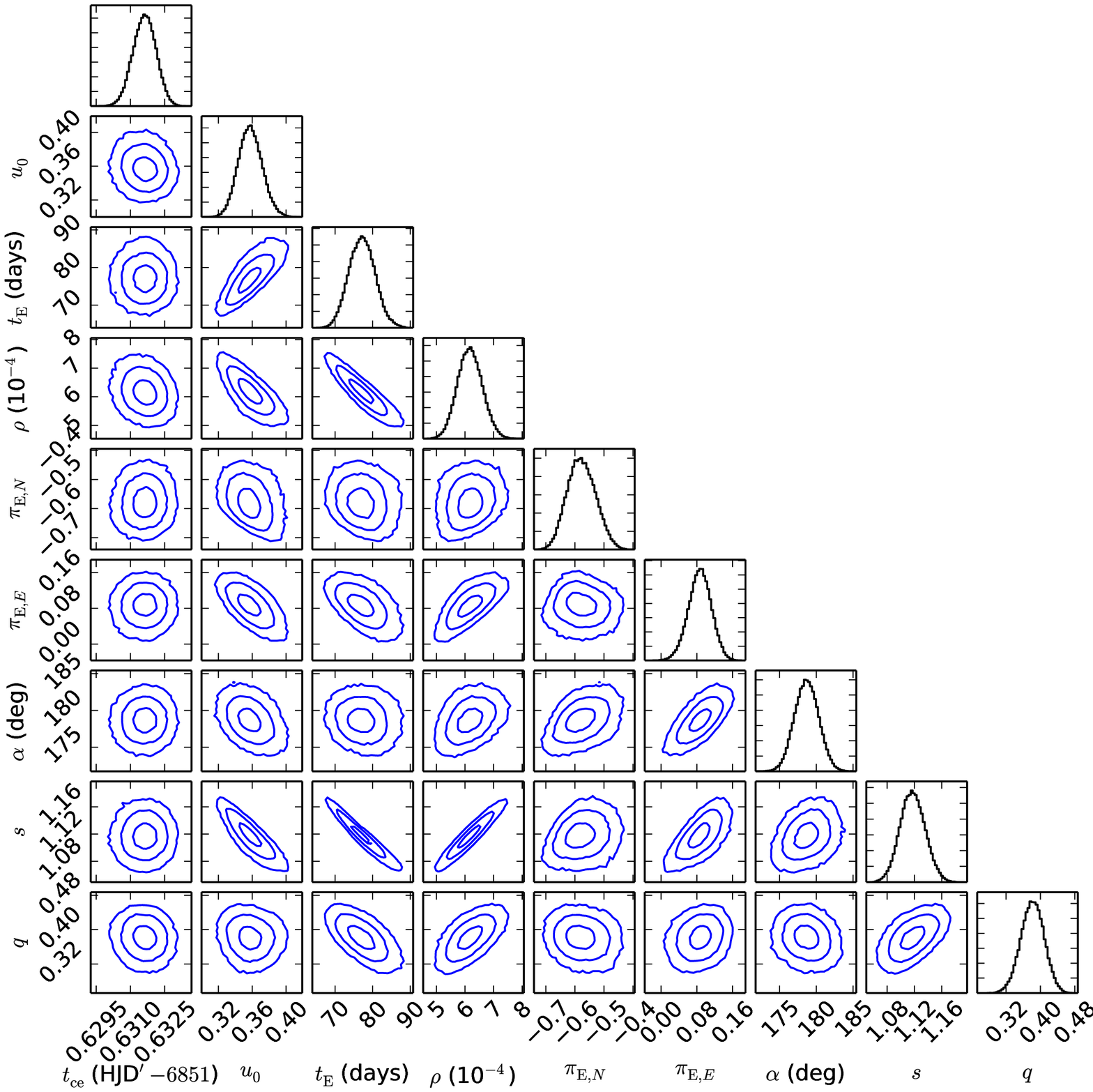}
\caption{Similar to Figure~\ref{fig:full-plus} but for the $(+,-)$ solution.}
\end{figure}

\clearpage
\begin{figure}
\centering
\plotone{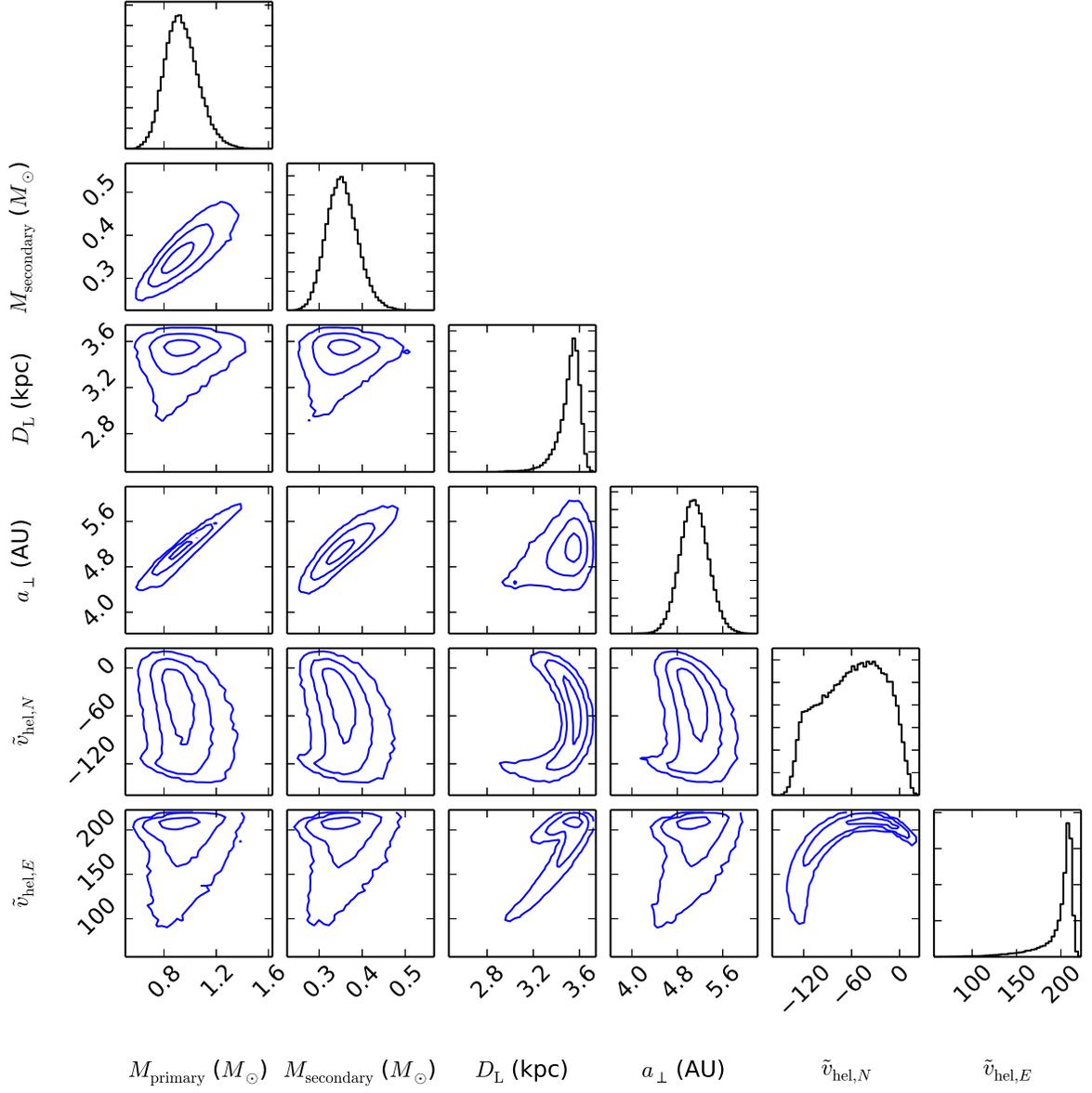}
\caption{The full triangle diagram of the derived physical parameters in the $(+,+)$ solution without orbital motion. Contours have the same meanings as in Figure~\ref{fig:full-plus}.
\label{fig:full-physical-plus}}
\end{figure}

\clearpage
\begin{figure}
\centering
\plotone{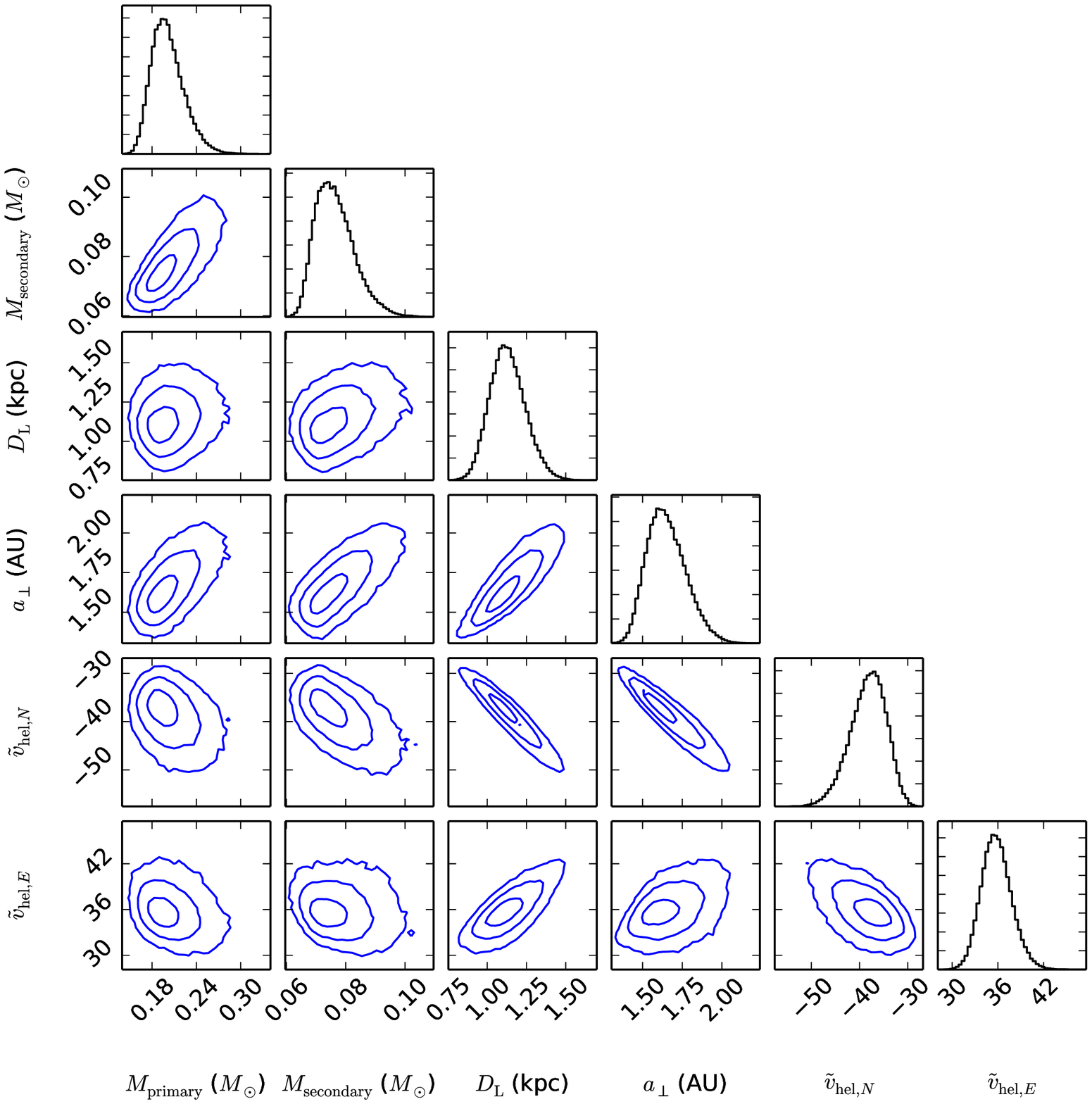}
\caption{Similar to Figure~\ref{fig:full-physical-plus} but for the $(+,-)$ solution.}
\end{figure}

\end{CJK*}
\end{document}